\documentclass[useAMS,usenatbib]{mn2e}
\usepackage{graphicx}
\usepackage{fancyref}
\usepackage{hyperref}
\hypersetup{colorlinks,breaklinks,
            urlcolor=[rgb]{0.2,0.2,0.2},
            linkcolor=[rgb]{0,0,0}, 
            citecolor=[rgb]{0,0.4,1.0}}

\def\aap{A\&A}
\def\apjl{ApJ}
\def\apj{ApJ}
\def\apjs{ApJS}
\def\aj{AJ}
\def\mnras{MNRAS}

\def\pasp{PASP}

\def\araa{ARA\&A}
\newcommand{\tar}{GALEX J1717+6757}
\newcommand{\msun}{$M_{\odot}$}
\newcommand{\rsun}{$R_{\odot}$}

\newcommand{\teff}{$T_{\rm eff}$}
\newcommand{\logg}{$\log{g}$}

\newcommand{\kms}{km~s$^{-1}$}
\newcommand{\gs}{g~s$^{-1}$}

\title[Far-UV abundances of an extremely low-mass white dwarf]{Heavy metals in a light white dwarf: Abundances of the metal-rich, extremely low-mass GALEX J1717+6757}

\author[Hermes et al.]{J.~J.~Hermes,$^{1}$\thanks{j.j.hermes@warwick.ac.uk}
B.~T.~G\"{a}nsicke,$^{1}$
D.~Koester,$^{2}$
M.~C.~P.~Bours$^{1}$,
D.~M.~Townsley,$^{3}$
\newauthor
J.~Farihi$^{4}$,
T.~R.~Marsh$^{1}$,
Stuart~Littlefair$^{5}$,
V.~S.~Dhillon$^{5}$,
A.~Gianninas$^{6}$,
\newauthor
E.~Breedt$^{1}$,
and R.~Raddi$^{1}$
\\
$^{1}$Department of Physics, University of Warwick, Coventry\,-\,CV4~7AL, UK\\
$^{2}$Institut f\"{u}r Theoretische Physik und Astrophysik, University of Kiel, Kiel\,-\,D-24098, Germany\\
$^{3}$Department of Physics and Astronomy, The University of Alabama, Tuscaloosa, AL\,-\,35487, USA\\
$^{4}$Department of Physics and Astronomy, University College London, Gower Street London\,-\,WC1E 6BT, UK\\
$^{5}$Department of Physics and Astronomy, University of Sheffield, Sheffield\,-\,S3 7RH, UK\\
$^{6}$Homer L. Dodge Department of Physics and Astronomy, University of Oklahoma, 440 W. Brooks St., Norman, OK\,-\,73019, USA}

\begin{document}

\label{firstpage}

\maketitle

\label{firstpage}

\begin{abstract}
Using the {\em Hubble Space Telescope}, we detail the first abundance analysis enabled by far-ultraviolet spectroscopy of a low-mass ($\simeq0.19$~\msun) white dwarf (WD), \tar, which is in a 5.9-hr binary with a fainter, more-massive companion. We see absorption from nine metals, including roughly solar abundances of Ca, Fe, Ti, and P. We detect a significantly sub-solar abundance of C, and put upper limits on N and O that are also markedly sub-solar. Updated diffusion calculations indicate that all metals should settle out of the atmosphere of this $14\,900$\,K, \logg\ $=5.67$ WD in the absence of radiative forces in less than 20 yr, orders of magnitude faster than the cooling age of hundreds of Myr. We demonstrate that ongoing accretion of rocky material that is often the cause of atmospheric metals in isolated, more massive WDs is unlikely to explain the observed abundances in \tar. Using new radiative levitation calculations, we determine that radiative forces can counteract diffusion and support many but not all of the elements present in the atmosphere of this WD; radiative levitation cannot, on its own, explain all of the observed abundance patterns, and additional mechanisms such as rotational mixing may be required. Finally, we detect both primary and secondary eclipses using ULTRACAM high-speed photometry, which we use to constrain the low-mass WD radius and rotation rate as well as update the ephemeris from the discovery observations of this WD+WD binary.

\end{abstract}


\section{INTRODUCTION}

The non-degenerate outer layers of white dwarf (WD) stars typically exhibit arresting simplicity: atmospheres that are featureless save for broad absorption lines of either hydrogen or helium. Most WDs are endowed with high surface gravities (\logg\ $\approx 8.0$), so metals quickly diffuse out of the photosphere on very short timescales, usually days to weeks for WDs which have radiative, hydrogen-dominated atmospheres \citep{Vauclair79,Paquette86,Koester09}. However, a number of WDs with impure atmospheres containing not only H or He but also heavier elements such as calcium, silicon, and iron have been known for nearly a century (e.g., \citealt{vanMaanen17,Weidemann60,Lacombe83,Holberg97}). These atmospheric metals require either an active process to counteract gravitational settling or an active replenishment mechanism.

\begin{figure*}
\centering{\includegraphics[width=0.925\textwidth]{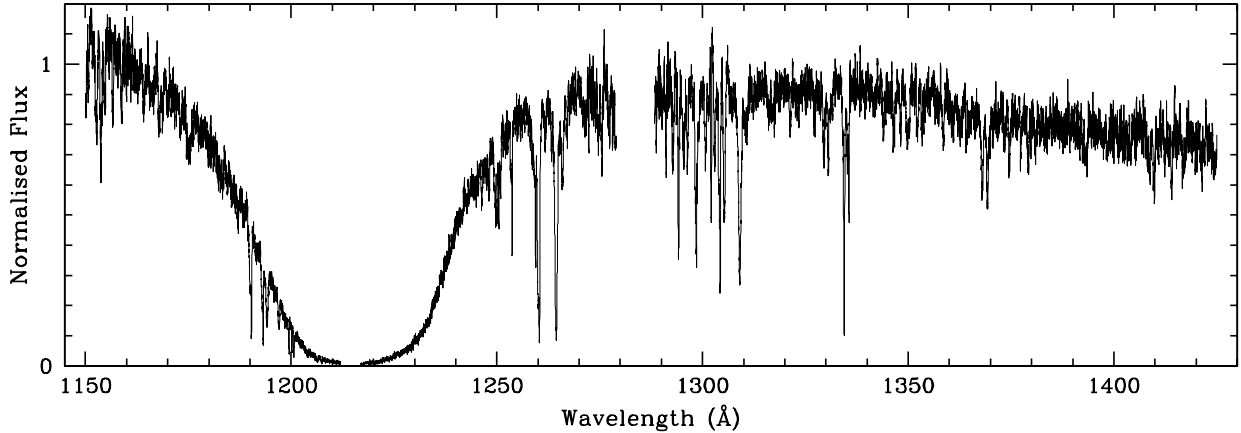}}
\caption{Our COS spectrum of \tar, smoothed by a 5-pixel Gaussian. The broad Ly-$\alpha$ feature dominates  (a geocoronal central component of Ly-$\alpha$ has been removed), but considerable metal absorption is present. Fig.~\ref{fig:speczoom} details the model fits to these metals. \label{fig:fullspectrum}}
\end{figure*}

A subset of the population of hydrogen-dominated (DA) WDs shows a remarkably high rate of detected atmospheric metals: the extremely low-mass (ELM, $<0.30$\,\msun) WDs, especially those with very low surface gravity, \logg\ $<6.0$. All of the lowest-gravity ELM WDs discovered to date show at least the Ca\,{\sc ii} K line in their spectra, and occasionally other metals \citep{Kaplan13}. Among the more than two dozen ELM WDs known to show Ca, there are three that exhibit additional metals in optical observations: \tar\ \citep{VK11}, PSR J1816+4510 \citep{Kaplan13}, and SDSS J0745+1949 \citep{Gianninas14}.

ELM WDs cannot have evolved from an isolated progenitor within the age of the Universe, and are by necessity the product of enhanced mass loss owing to a binary companion, probably during a common-envelope phase \citep{Iben84,Iben85,Marsh95,RM11}. Diffusion times for metals increase substantially with decreasing surface gravity, but in all ELM WDs the cooling ages still exceed the diffusion timescales in the absence of supportive forces, including radiative levitation, by several orders of magnitude.

Metals are also observed in the atmospheres of some DAs near the canonical mean mass for WDs, roughly 0.6\,\msun\ (e.g., \citealt{Bergeron92,Falcon10,Tremblay11,Kleinman13}). Roughly $20-50$ per\,cent of cool, average-mass WDs show evidence of atmospheric metals, usually through Ca, which is often the easiest metal to detect at optical wavelengths \citep{Zuckerman03,Zuckerman10,Koester14}.

For average-mass WDs hotter than roughly $20\,000$\,K, radiative levitation can support some amount of heavy metals out to the photosphere (e.g., \citealt{Chayer95,Barstow14,Chayer14}). However, the majority of heavy metal abundances observed in cool WDs are best explained by active accretion of debris from tidally disrupted asteroids or minor planetary bodies. This opens the exciting possibility of constraining the bulk composition of extrasolar planetary material (e.g., \citealt{Jura03,Zuckerman07,Dufour12,Gaensicke12}). In fact, polluted WD abundances have indicated that some exoplanetary debris is both rocky and water-rich \citep{Farihi13}.

Aside from rapid settling time arguments, the best evidence that metal-rich WDs undergo accretion from exoplanetary debris that survived the late stages of stellar evolution come from observed gaseous metal disks and infrared excesses around many metal-polluted WDs, which provide evidence that circumstellar disks exist and provide a reservoir of debris (e.g., \citealt{Becklin05,Gaensicke06}). While not all metal-rich WDs are observed with debris disks, all infrared excesses are in systems with metal-polluted WDs \citep{vonHippel07,Farihi12}.

The ubiquity of metals in the lowest-gravity WDs, which more frequently show atmospheric metals than average-mass WDs, provides an opportunity to constrain any additional effects counteracting diffusion in these low-mass WDs. Thus, we have carried out a follow-up investigation of one such star, the relatively bright ($V=13.7$\,mag) \tar. The system was discovered by \citet{VK11}, who solved the orbital parameters of this double-degenerate system ($P=5.9$~hr) and found a bright, low-mass ($\simeq0.19$~\msun) WD orbited by a fainter companion. They detected shallow eclipses, which requires the system to be at high inclination, and thus $M_2 \simeq 0.90$~\msun. The identification of photospheric Si, Ca, and Fe, in phase with the Balmer lines, allowed \citet{VK11} to conduct the first abundance analysis of an ELM WD.

We have expanded on this initial abundance analysis by observing \tar\ using the Cosmic Origin Spectrograph (COS) on board the {\em Hubble Space Telescope} ({\em HST}), providing the first far-ultraviolet spectrum of an ELM WD. We present our observations, reductions, and derived metal abundances in Section~2, where we also detail new high-speed photometry of this eclipsing binary. We discuss the possible origins of these metals in ELM WDs in Section~3 and conclude in Section~4.

\section{OBSERVATIONS AND ANALYSIS}

\subsection{Far-ultraviolet spectrum}

We observed GALEX J171708.5+675712 (hereafter \tar) as part of a far-ultraviolet spectroscopic survey of the frequency and chemical composition of rocky planetary debris around hydrogen-dominated WDs, {\em HST} Cycle~19 programme~12474. Fig.~\ref{fig:fullspectrum} details our 800\,s COS spectrum on 2011~December~26 using the G130M grating \citep{Green12}, where we used a central wavelength of $1291$\,\AA\ and covered the wavelength range $1130-1435$\,\AA. There is a small gap between $1278-1288$\,\AA\ from the two segmented detectors. The data were retrieved from the {\em HST} archive and reduced and processed using {\sc calcos 2.15.6}.

The broadest absorption feature in the spectrum corresponds to the Stark-broadened Ly-$\alpha$ line, but a host of narrow absorption features are evident, that, as we later demonstrate, arise from a range of photospheric metals and are broadened by rotation. Using a relatively line-free region from $1320-1330$\,\AA, we calculate a mean signal-to-noise ratio (S/N) $\approx15$. Our resolution ranges from $R\approx15\,000-20\,000$ between 1150\,\AA\ and 1435\,\AA, respectively.

\subsection{Metal abundance determinations}

We began by calculating the atmospheric parameters from the new far-ultraviolet observations. We fit the spectrum with the latest DA models of \citet{Koester10}, and find \teff\ = $14\,110$~K, \logg\ $= 5.47$, but with rather large uncertainties from the {\em HST} data alone. Therefore we adopt the determinations of \citet{VK11}, \teff\ $=14\,900 \pm 200$\,K and \logg\ $= 5.67 \pm 0.05$, which were found by fitting an optical spectrum and match well the available ultraviolet to infrared photometry.

The determination of abundances and upper limits proceeded in the same way as described in detail in \citet{Gaensicke12}. In short, by visual inspection we identified the elements with strong resonance lines in our wavelength range (see Table~2 in \citealt{Gaensicke12}) and modified the abundances of the elements until a reasonable fit to the spectrum was obtained. The abundances were then increased and decreased by 0.3 and 0.5 dex, until the fit was clearly worse. This informed the estimates of the abundance uncertainties.

\begin{table}
 \centering
  \caption{Metal abundances, diffusion times, and element diffusion fluxes in \tar.}
  \label{tab:abundances}
  \begin{tabular}{@{}lccr@{}}
  \hline
Z  & log(Z/H)   & log($\tau_{\rm diff}$) & log(Diff. Flux) \\
    &            & (yr)  & (\gs)            \\
 \hline
C  & $-7.2$(0.5) & 0.951 & 4.665 \\
N  & $<-6.3$     & 0.741 & $<5.824$ \\
O  & $<-5.9$     & 0.579 & $<6.434$ \\
Mg & $<-5.0$     & 1.309 & $<6.852$ \\
Al & $-6.7$(0.5) & 1.116 & 5.387 \\
Si & $-5.7$(0.3) & 1.043 & 6.472 \\
P  & $-7.0$(0.5) & 0.796 & 5.456 \\
S  & $-6.0$(0.5) & 0.663 & 6.602 \\
Ca & $-5.2$(0.5) & 1.132 & 7.050 \\
Sc & $<-6.0$     & 1.077 & $<6.353$ \\
Ti & $-6.0$(0.5) & 1.045 & 6.413 \\
Cr & $-6.0$(0.5) & 0.971 & 6.523 \\
Fe & $-5.3$(0.5) & 0.885 & 7.340 \\
Ni & $<-6.0$     & 0.780 & $<6.765$ \\
\hline
\end{tabular}
\end{table}

\begin{figure}
\centering{\includegraphics[width=0.991\columnwidth]{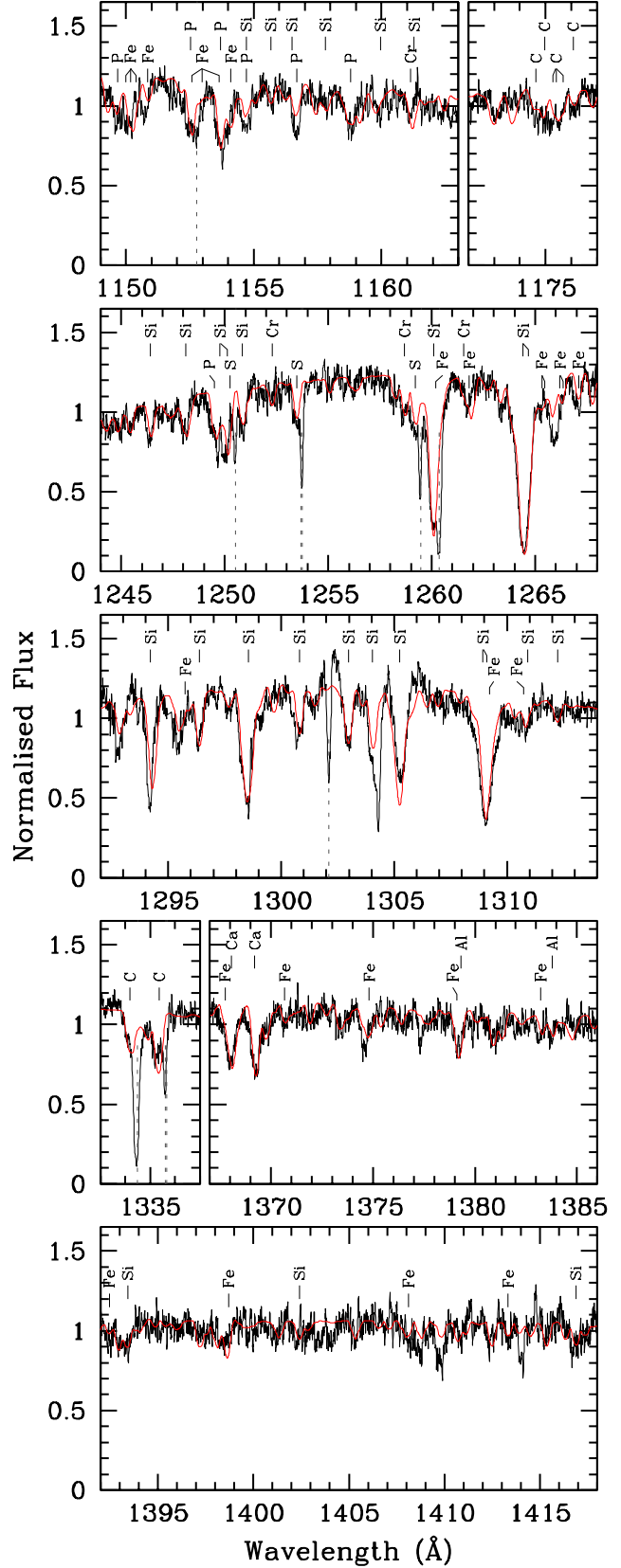}}
\caption{The normalized far-ultraviolet spectrum of \tar\ is shown in black, along with our best-fit atmosphere model in red, which includes hydrogen and nine heavier metals. Interstellar absorption features are marked with vertical dark grey dashed lines and are not included in the fit. \label{fig:speczoom}}
\end{figure}

\begin{figure*}
\centering{\includegraphics[width=0.9995\textwidth]{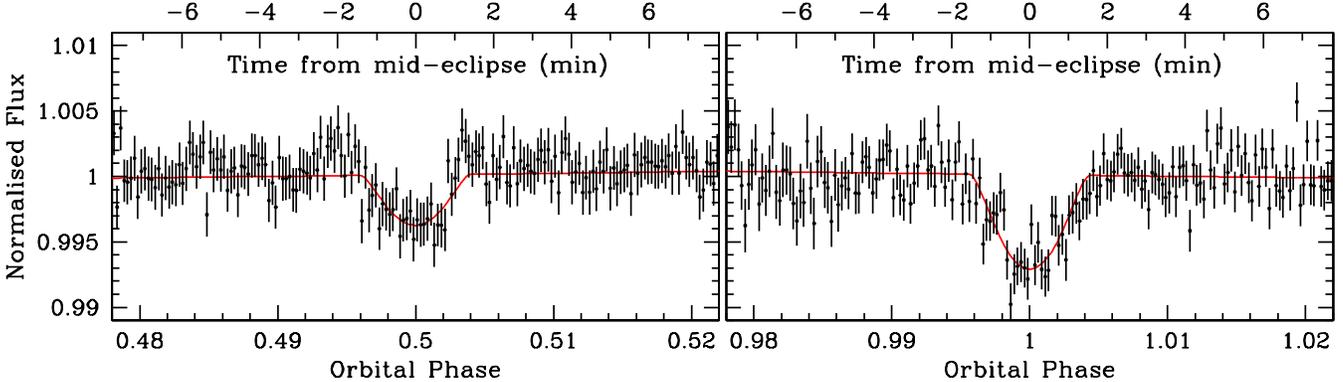}}
\caption{Phase-folded, SDSS-{\em g} light curves of \tar\ obtained using ULTRACAM on the 4.2\,m WHT. Our observations cover two primary eclipses of the higher-mass companion transiting the ELM WD at phase~0.5 (left panel) and three secondary eclipses of the ELM WD transiting the fainter, higher-mass companion at phase~1 (right panel). The solid red curve displays our best-fit model.
\label{fig:lightcurve}}
\end{figure*}

We detail our abundance determinations in Table~\ref{tab:abundances}. Relative to H, we measure the photospheric abundances of nine elements: C, Al, Si, P, S, Ca, Ti, Cr, and Fe. Additionally, we place upper limits on the presence of N, O, Mg, Sc, and Ni, which have moderately strong transitions in the wavelength range of our COS observations. Fig.~\ref{fig:speczoom} shows several regions in the spectrum of \tar\ with the most prominent lines; we over-plot the best-fit model.

We detect interstellar lines of C, Si, P, S, and N, marked with grey dashed lines in Fig.~\ref{fig:speczoom}. These interstellar lines are blue shifted relative to the photospheric lines by $v = 64 \pm 15$~\kms; our COS observations of \tar\ occur with a radial velocity of $v=-77\pm15$~\kms, which agrees within $2\sigma$ with the uncertainties with the predicted, phased radial velocity of \citet{VK11}. We also note the presence of Earth airglow emission, which contaminates the region from $1301-1307$\,\AA\ with geocoronal lines of O\,{\sc i}. However, there is an additional relatively uncontaminated line of O\,{\sc i} at 1152.15\,\AA\ which we use to constrain the O abundance.

The Si, Ca, and Fe abundances derived from the optical spectrum of \citet{VK11} agree with our determinations, within the uncertainties.

\subsection{Light curve and stellar rotation}
\label{sec:lc}

Our abundance uncertainties are rather large and similar for most elements because all lines are strongly broadened, very likely by rotation. To determine the rotational velocity we convolved the final model with Gaussian profiles corresponding to 10, 30, 50, 70, and 100~\kms. Another convolution with a Gaussian of 0.1~\AA\ was then applied to take into account the spectral resolution of the instrument. The best fit was again obtained by visual inspection and returned with $v \sin i = 50^{+30}_{-20}$ \kms.

In order to better constrain the physical parameters of this eclipsing binary, we have obtained follow-up high-speed photometry using ULTRACAM \citep{Dhillon07} mounted on the 4.2\,m William Herschel Telescope (WHT) on La Palma. These data were reduced using the ULTRACAM pipeline software, with standard bias correction, flat-fielding, and aperture photometry reductions.

These follow-up observations were simultaneously obtained with SDSS-{\em u},{\em g},{\em r} filters on the nights of 2011~Aug~20 (21.6\,min using 0.8\,s exposures around phase~1), 2011~Aug~21 (25.6\,min using 0.5\,s exposures around phase~0.5), 2012~Sep~08 (38.9\,min using 1.5\,s exposures around phase~1), and 2014~Mar~29 (39.6\,min using 0.5\,s exposures around phase~0.5), as well as through SDSS-{\em u},{\em g},{\em i} filters on 2013~Jul~31 (19.3\,min using 1.3\,s exposures around phase~1).

\begin{table}
 \centering
  \caption{New mid-eclipse times of \tar.}
  \label{tab:eclipse}
  \begin{tabular}{@{}lr@{}}
  \hline
Cycle  &  BMJD(TDB) \\
 \hline
620    & $55\,794.035261(66)$ \\
623.5  & $55\,794.896646(70)$ \\
2184   & $56\,178.991009(53)$ \\
3508   & $56\,504.874289(41)$ \\
4614.5 & $56\,777.222879(59)$ \\
\hline
\end{tabular}
\end{table}

We recover the secondary eclipses of the ELM WD transiting the higher-mass companion at phase~1 first reported by \citet{VK11}, and for the first time detect primary eclipses of the more massive WD passing in front of the ELM WD at phase~0.5\footnote{We retain the convention of \citet{VK11} of defining the less-massive WD as the primary, since it is the brighter WD.}. The eclipses in all three filters are of equivalent depth and mid-eclipse times, within the uncertainties, so we display in Fig.~\ref{fig:lightcurve} our highest-quality light curve through the SDSS-{\em g} filter, binned into 4000 orbital phase bins. Plotted to guide the eye in Fig.~\ref{fig:lightcurve} is our best-fit model reproducing the shallow (0.3 per\,cent) primary eclipses and the slightly deeper (0.7 per\,cent) secondary eclipses. We computed this model using a light curve code for binary systems containing at least one WD (see \citealt{Copperwheat10}) using the $g$-band, Claret four-parameter limb-darkening coefficients of \citet{Gianninas13}.

\begin{figure*}
\begin{center}
 \includegraphics[width=0.92\textwidth]{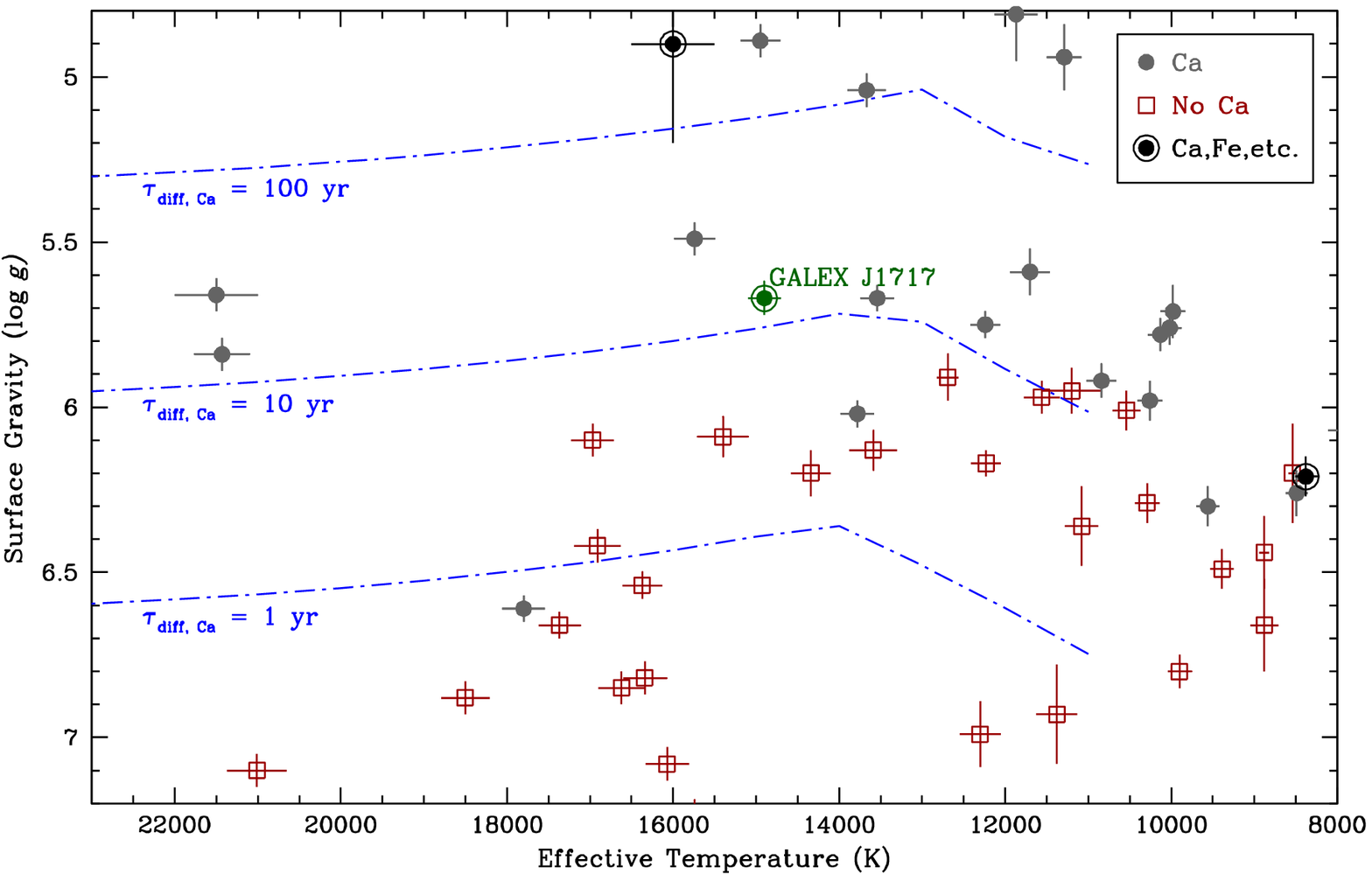}
 \caption{The \logg-\teff\ diagram for low-mass WDs. Grey filled circles are ELM WDs with Ca detected in their optical spectra, and red open squares are ELM WDs that do not show Ca absorption lines. All ELM WDs with \logg\ $<5.9$ show Ca. We encircle the three objects with additional metals beyond Ca; these are, in order of decreasing effective temperature, PSR~J1816+4510 \citep{Kaplan13}, \tar\ \citep{VK11}, and SDSS~J0745+1949 \citep{Gianninas14}. The other points come from visual inspection of the spectra from the ELM Survey \citep{BrownELMv}. We plot contours of constant Ca diffusion settling times, from $1-100$ yr, as blue dot-dashed lines; this is the timescale for Ca to diffuse out of the photosphere without additional support.}
 \label{fig:loggteff}
 \end{center}
\end{figure*}

In light of the shallow eclipses and the marginal S/N of our data, we hold fixed $T_1 = 14\,900$~K and assume $R_2=0.0093$~\rsun, given the expected radius of a 0.9~\msun\ WD \citep{Fontaine01}\footnote{See: \href{http://www.astro.umontreal.ca/~bergeron/CoolingModels}{http://www.astro.umontreal.ca/$_{\widetilde{~}}$bergeron/CoolingModels}}.
Fitting for the inclination, $i$, the primary radius, $R_1/a$, the secondary temperature, $T_2$, and the zero-point of the ephemeris, $T_0$, we find $i=86.9\pm0.4$~deg, $R_1=0.093\pm0.013$~\rsun, and $T_2 = 15\,500\pm1800$~K. The uncertainties quoted are 3$\sigma$ and arise from fitting all five eclipses simultaneously using a Markov chain Monte Carlo analysis. We also update the ephemeris to
$$ \rm{BMJD}_{\rm TDB} = 55641.431322(65) + 0.246135373(17) \; E $$
consistent with the ephemeris of \citet{VK11} but with a more precise period given our longer baseline.

Our far-UV spectrum is best fit with a one-WD model, which strongly suggests that $T_2<25\,000$\,K; otherwise, the more massive secondary would be detected in the Ly-$\alpha$ core. Our estimate of $T_2$ from the ratio of the eclipse depths is consistent with this constraint.

\citet{VK11} suggested that gravitational deflection might explain the shallowness of primary eclipses at phase~0.5 \citep{Marsh01}. If we do not include gravitational lensing in our best-fit model, the depth of the primary eclipse changes from 0.29 per\,cent to 0.38 per\,cent. Our light curve fits including gravitational lensing provide a marginally better fit than those without: The fits have a $\chi^2/{\rm dof}=25\,756/9156$ including the effect of gravitational lensing and a $\chi^2/{\rm dof}=25\,774/9156$ without lensing (we have rescaled the uncertainties in Fig.~\ref{fig:lightcurve} such that $\chi^2/{\rm dof}=1.0$). For the latter model, the lack of gravitational lensing was compensated for by requiring the more massive companion to be slightly hotter, and this best-fit model returned $i=86.6\pm0.4$~deg, $R_1=0.101\pm0.014$~\rsun, and $T_2 = 17\,900\pm1900$~K. In either case, $L_2< 1.5$ per\,cent of the total system luminosity.

Additionally, we see marginal evidence for the mid-eclipse times of the primary eclipses at phase~0.5 occurring slightly sooner than those at phase~1.0 (by $12.2\pm6.6$~s). We expect a slight light-travel-time delay between the mid-eclipse times of the primary and secondary eclipses as a result of the large mass ratio of the two WDs. In the case that $M_1<M_2$, \citet{Kaplan10} show the R\o mer delay goes as
$$\delta_R = \frac{P_{\rm orb} K_1}{\pi c} (1-M_1/M_2)$$
so in \tar\ we expect the eclipses centered at phase~0.5 to occur $5.1\pm0.3$~s sooner than those at phase~1.0. Further eclipse monitoring can better constrain any small eccentricity of the orbit.

Given the effective temperature and surface gravity, the most recent He-core WD models of \citet{Althaus13} suggest an overall mass of $M_1\simeq0.19$\,\msun\ and a radius of $0.105$\,\rsun\ for \tar, which is only slightly larger than what we find empirically from the eclipses. Using our updated radius, $R_1=0.093\pm0.013$\,\rsun, the observed rotational velocity corresponds to a rotation rate of $2.3^{+2.0}_{-1.0}$~hr, close to the orbital period, albeit marginally (but not significantly) faster.

\section{DISCUSSION}

\begin{figure*}
\begin{center}
 \includegraphics[width=0.99\textwidth]{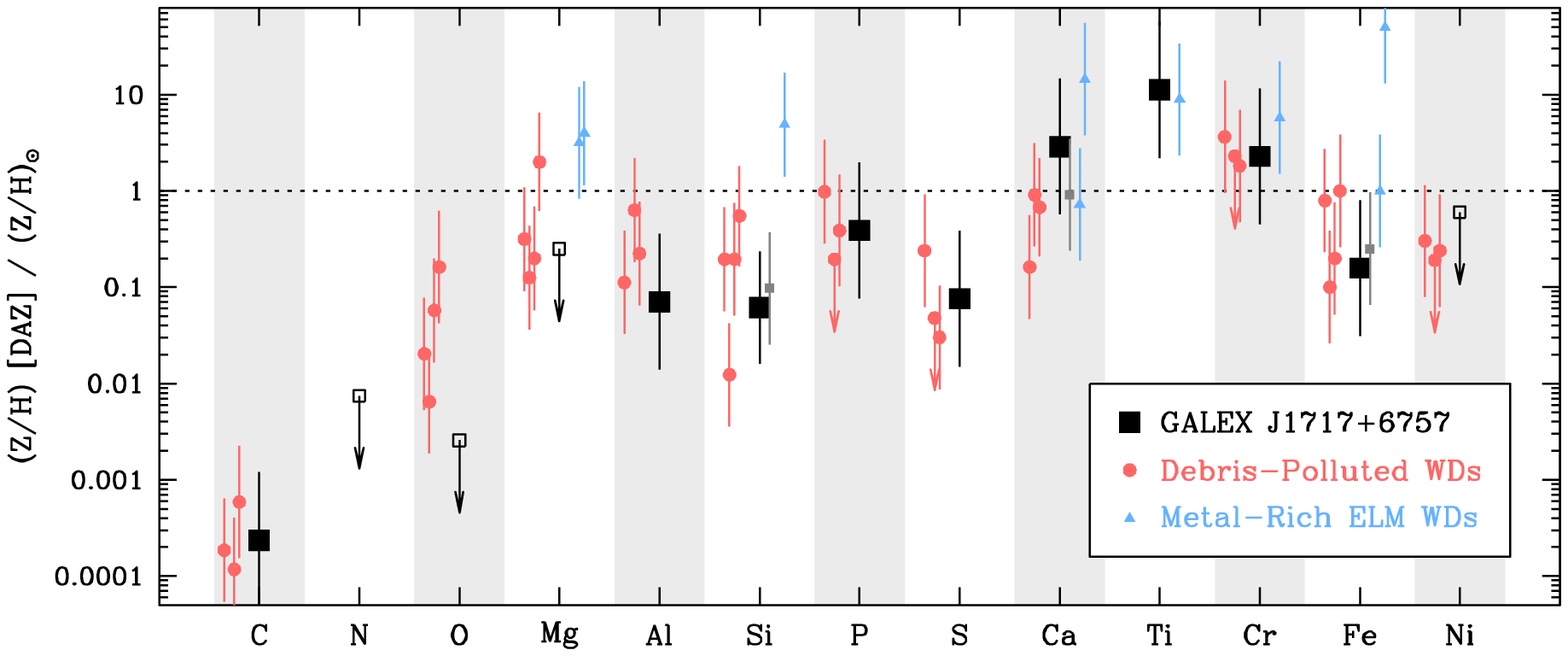}
 \caption{Element abundances in \tar, detailed in Table~\ref{tab:abundances}. Our measured abundances are displayed as large black squares relative to solar \citep{Asplund09}, marked by the dotted line, and upper limits are shown as open black squares. The abundance determinations from the optical spectrum of \tar\ by \citet{VK11} are included as smaller dark gray squares. Note that this figure spans more than six orders of magnitude. We include, as light red points left of our target, abundances from four average-mass WDs that have debris disks and are polluted from exo-terrestrial planetesimals; these WDs were analysed in an identical way by \citet{Gaensicke12}. We also include abundances for the metal-rich low-mass WDs PSR~J1816+4510 \citep{Kaplan13} and SDSS~J0745+1949 \citep{Gianninas14} as light blue points.}
 \label{fig:abundances}
 \end{center}
\end{figure*}

\subsection{Metals in low-gravity WDs}

\citet{Kaplan13} featured one of the first illustrations of the ubiquity of atmospheric metals (usually by the Ca\,{\sc ii} K line) detectable in the optical spectra of ELM WDs. They showed that all ELM WDs with high-quality spectra and with \logg\ $<5.6$ have atmospheric Ca. Fig.~\ref{fig:loggteff} updates this result using the most recent results from the ELM Survey (see \citealt{BrownELMv} and references therein).

We plot in this \logg--\teff\ diagram only ELM WDs with sufficiently high S/N to detect the Ca\,{\sc ii} K line, using updated atmospheric parameters (Gianninas et al. 2014, submitted). In some cases this Ca may be interstellar, and confirmation of a photospheric origin requires the Ca and Balmer lines to be detected at the same radial velocity. However, for all ELM WDs bright enough to detect Ca in a single exposure, the Ca\,{\sc ii} K line phases with the Balmer lines (W. Brown 2014, private communication). A clear delineation in the population of ELM WDs with and without photospheric Ca can be drawn for the targets with \logg\ $<5.9$. There is no appreciable trend in detectable Ca with effective temperature, nor with binary orbital period.

Diffusion rates for metals in WDs have never been previously calculated for \logg\ $<7.0$, so in order to explore the rates for these low-mass WDs we have extended the diffusion timescale calculations of \citet{Koester09} down to \logg\ $= 5.0$. These diffusion times only include gravitational acceleration but no radiative support. We plot in Fig.~\ref{fig:loggteff} contours of constant diffusion timescales, in order to ascertain if diffusion simply acts sufficiently slow to maintain metals in the atmospheres of the lowest-gravity WDs. Our new calculations demonstrate that diffusion rates are several orders of magnitude shorter than evolutionary timescales. For ELM WDs on the terminal cooling track (those no longer experiencing CNO flashes, see Section~\ref{sec:CNO}), the cooling ages are more than 100 Myr for all objects in Fig.~\ref{fig:loggteff}.

We have also calculated the specific diffusion times for each element given the atmospheric parameters for \tar, and list the values in Table~\ref{tab:abundances}. These settling times, ranging from between 4\,yr to 20\,yr, are vastly shorter than the predicted cooling age of $410\pm130$\,Myr \citep{Althaus13}.

\subsection{Source of the metals}

Given the discrepancy in gravitational settling versus evolutionary times, the metals in ELM WDs require either an active replenishment mechanism, such as external accretion, or an internal physical process counteracting diffusion, such as radiative levitation or rotational mixing. Fig.~\ref{fig:loggteff} demonstrates a strong anti-correlation between the detection of metals and the WD surface gravity. While accretion can explain the presence of photospheric metals, it appears highly unlikely that the existence of the necessary reservoir of polluting material also anti-correlates with the WD surface gravity.

\subsubsection{Problems with pollution from ongoing accretion}

Among the strong evidence that metal abundances observed in average-mass WDs arise from the accretion of planetary debris are the low C abundances, consistent with Solar System meteorites \citep{Jura06}. In order to place our metal-rich ELM WD in this context, we show in Fig.~\ref{fig:abundances} the abundances of all elements detected (and some upper limits) in \tar\ relative to solar \citep{Asplund09}.

Notably, \tar\ is quite deficient in C, similar to four average-mass, metal-polluted WDs that are actively accreting rocky planetesimals from simultaneously observed debris disks \citep{Gaensicke12}. However, this ELM WD has O and Mg abundances atypical of debris-polluted WDs.

Table~\ref{tab:abundances} details the diffusion fluxes calculated for the observed abundances in \tar. In the absence of competing forces such as radiative levitation, this would correspond to a total accretion rate of roughly $6.5\times10^7$\,\gs, close to a typical time-averaged accretion rate for known metal-rich WDs \citep{Farihi09,Girven12,Koester14}. However, these diffusion fluxes likely rule out the accretion of rocky material as the source of the observed metals, since we would expect the heavy elements to be delivered through metal oxides (e.g., \citealt{Xu13,Farihi13}). Following \citet{Klein10}, we can calculate the amount of O contained in the most common metal oxides, which we show in Table~\ref{tab:oxygen}.

\begin{table}
 \centering
  \caption{Expected oxygen budget assuming accretion of metal oxides in \tar.}
  \label{tab:oxygen}
  \begin{tabular}{@{}lr@{}}
  \hline
Oxygen carrier  &  Expected oxygen mass flux \\
 \hline
CaO             & $4.5\times10^6$~\gs\ \\
${\rm SiO_2}$   & $3.4\times10^6$~\gs\ \\
${\rm Al_2O_3}$ & $2.2\times10^5$~\gs\ \\
${\rm TiO_2}$   & $3.4\times10^6$~\gs\ \\
{\bf Total expected O mass flux}  & {\bf ${\bf >11.4\times10^6}$~\gs} \\
{\bf Observed O mass flux}  & {\bf ${\bf <2.7\times10^6}$~\gs} \\
\hline
\end{tabular}
\end{table}

From the observed Ca, Si, Al, and Ti diffusion fluxes alone, we would expect $>1.14\times10^7$\,\gs of O delivered through ${\rm SiO_2}$, CaO, ${\rm Al_2O_3}$, and ${\rm TiO_2}$. This is a lower limit, as it does not include FeO or ${\rm Fe_2O_3}$ (since Fe could be delivered in metallic form) or volatiles like ${\rm H_2O}$, and it is a factor of four higher than the upper limit on the diffusion flux for O we infer from our observations, $<2.7\times10^6$\,\gs. Even if we assume the lowest diffusion fluxes of Ca, Si, Al, and Ti that are consistent with our abundance uncertainties, we would still expect an O diffusion flux $>4.1\times10^6$\,\gs. Given the observed heavy element abundances, O would have been detected in \tar\ if ongoing accretion of rocky debris material were responsible for its atmospheric metals, as observed in more massive metal-polluted WDs.

The hypothesis of recent external accretion onto \tar\ faces another major challenge: the star is a 5.9-hr binary with a 0.9 \msun\ companion. The sheer proximity of these two objects ($a\simeq1.7$ \rsun) makes accretion onto the ELM WD from a circumbinary disk dynamically difficult, if not impossible, especially if the potentially accreted object has completed a few stable orbits \citep{Veras14}. Still, it is worth noting that smoothed particle hydrodynamics simulations have shown it possible to supply a small amount of material from a circumbinary disk onto the central binary given the right conditions \citep{Artymowicz96}.

In summary, we consider it unlikely that ongoing accretion of rocky material is responsible for the metals present in the atmosphere of \tar.

\subsubsection{Radiative acceleration}

Fig.~\ref{fig:loggteff} demonstrates a clear trend in the frequency of Ca detection with the lowest-gravity WDs that is consistent with an internal stellar process counteracting diffusion that increases in strength with decreasing surface gravity. Radiative levitation is one such mechanism. There is growing evidence that radiative forces play an important role in the metal abundance patterns of many WDs, especially for light species such as C, Si, and Al \citep{Dupuis10}, and levitation has a direct impact on derived accretion rates for WDs polluted by exoterrestrial debris \citep{CD10,Chayer14}.

Unfortunately, there have been few investigations into the full range of situations in which only radiative support can effectively maintain metals in WD atmospheres, and none for ELM WDs. The calculations of \citet{Chayer95} predict minimal radiative support in the photosphere for any element below $20\,000$\,K for the lowest-gravity WD calculated, \logg\ $=7.0$. This surface gravity is more than an order of magnitude higher than the regime of the ELM WDs, including \tar.

We have implemented a new code with radiative levitation \citep{Koester14}, and tested for most of the metals observed in \tar: C, N, O, Si, P, Ca, and Fe. We predict minimal support for N and O, compatible with the upper limits observed\footnote{The previous discussion of ongoing accretion of rocky material took the diffusion fluxes at face value, but radiative levitation can alter these fluxes.}. The calculations predict strong radiative support for Fe and moderate support for Si and P, also fully consistent with the observed abundances. 

However, the radiative levitation predictions are not completely in line with our observations. We expect strong radiative support for C, predicting an abundance hundreds of times higher than seen in \tar. Additionally, we predict minimal radiative support for Ca, far less than required to explain the observed Ca abundance.

Thus, radiative levitation can account for many of the metals we see in \tar, but it cannot fully explain the observed abundance patterns. Additional support mechanisms are likely necessary to explain the atmospheric composition of the lowest-gravity WDs like \tar.

Horizontal branch stars provide useful context for radiative support of metals, since a similar, orders-of-magnitude discrepancy between the diffusion and evolutionary timescales initially plagued abundance studies of the hot subdwarf (sdB) stars. Diffusion alone cannot explain the observed high helium abundances, as well as maintaining sufficient iron in the upper atmosphere to drive pulsations in sdBs (e.g., \citealt{Charpinet96,Fontaine03,Jeffery06}). One explanation invokes radiative accelerations, perhaps driven by weak stellar winds \citep{Fontaine97} or turbulent mixing \citep{Michaud11,Hu11}. Hot subdwarfs have similar surface gravities as ELM WDs, but undergo core helium burning, which leads to typically hotter surfaces and thus harder radiation fields.

In addition, blue horizontal branch stars are evolved objects nearer in temperature to ELM WDs that also show abundance anomalies (e.g., \citealt{Behr03}). As with sdBs, theoretical predictions of strong radiative levitation can explain many of the discrepancies \citep{Michaud83}.

\subsubsection{Rotational mixing}

An additional support mechanism to consider is rotational mixing (such as meridional circulation or rotationally induced turbulence), which can work against diffusion in rapidly rotating stars (e.g., \citealt{Vauclair78,Michaud80,Talon06}).

For WDs, \citet{SW71} investigated the cases where rotation could counteract diffusion, and suggested that diffusion in WD atmospheres would be inhibited if $(v_{\rm rot} / v_{\rm b}) > 0.008$, where  $v_{\rm rot}$ and $v_{\rm b}$ are the WD rotation and breakup velocities, respectively. Given our observational results from Section~\ref{sec:lc}, we expect the ELM WD in \tar\ to have a relatively slow breakup velocity of $950\pm140$~\kms, so $(v_{\rm rot} / v_{\rm b}) = 0.05\pm0.03$. The breakup velocity drops rapidly with decreasing surface gravity, since $v_{\rm b} \propto \sqrt{g/R}$ and lower-gravity WDs have larger radii, so this ratio becomes much larger for the lowest-gravity WDs\footnote{Note that this criterion is not met for a typical, 0.6~\msun\ WD. We observe from asteroseismology that isolated WDs typically rotate at periods of roughly 1~d \citep{Kawaler04}, so for an average-mass, \logg\ $=8.0$ WD, $(v_{\rm rot} / v_{\rm b}) \sim 10^{-5}$.}.

Unfortunately we do not know rotation rates for the low-mass WDs in Fig.~\ref{fig:loggteff}, but in the absence of other information we can assume they are tidally locked and rotating at their orbital periods. With this assumption, we find that none of the systems with $v_{\rm rot} / v_{\rm b} < 0.004$ show atmospheric metals. However, several systems with $P_{\rm orb} < 2$~hr do not exhibit photospheric metals despite having $v_{\rm rot} / v_{\rm b} > 0.01$.

While this ratio may be an overly simplistic diagnostic, using it to assess relative trends may help constrain a physical mechanism responsible for possible rotational mixing.

\subsection{The underabundances of C, N, and O}
\label{sec:CNO}

Another curiosity in the atmospheric makeup of \tar\ is the abundances of the lightest metals: C, N, and O. As shown in Fig.~\ref{fig:abundances}, these species are all significantly underabundant relative to the solar value, by at least a factor of 100, with C more than 1000 times below the solar value.

Here again, comparison with the hot subdwarfs is useful. Detailed spectroscopic studies show that the majority of sdBs have sub-solar C abundances, down to $1/300$ the solar value; especially C deficient are the sdBs on the lower helium sequence, objects without core helium burning \citep{Geier13a}. Lower helium sequence sdBs are not clearly post-red-giant-branch stars \citep{Geier13b}, but they could represent progenitors of He-core, ELM WDs. O abundances in sdBs are also significantly sub-solar, down to $1/100$ solar, but N abundances are generally closer to the solar value. Subdwarf CNO abundances thus resemble in some aspect the ones we observe in \tar.

Stellar evolution models that include the effects of atomic diffusion and radiative acceleration can effectively explain the abundances in horizontal branch stars, including low CNO abundances observed in subdwarfs \citep{Michaud11}. Perhaps most relevant, calculations for sdBs have shown that stellar winds can lead to strong deficiencies (or enrichment) in the CNO elements, depending on how long diffusion and mass loss have occurred. Such history may have shaped the low CNO abundances observed in \tar, since the lowest-gravity WDs with \teff\ $>20\,000$ K are near the wind limit (see Fig. 8 of \citealt{Unglaub01}).

Although the main-sequence progenitor of \tar\ did not reach central temperatures sufficient to initiate CNO burning, it is possible that CNO processing occurred after this He-core WD emerged from its common-envelope phase. Diffusion-induced CNO flashes are expected from theoretical models of all but the least massive ($M\leq 0.18$\,\msun) ELM WDs, as the diffusive hydrogen tail reaches sufficiently deep in the WD to ignite CNO burning (e.g., \citealt{IT86,Driebe99,Panei07,Althaus13}).

These CNO flashes initiate considerable surface convection zones, which can effectively dredge up material, including metals, to the stellar photosphere (M. Montgomery 2013, private communication). It is possible, then, that these flashing episodes bear responsibility for at least some of the high metal abundances observed in ELM WDs, and for explaining the significantly subsolar C abundance we see in \tar. Although we expect strong support for C from radiative levitation, perhaps the low C abundance observed reflects a history of depletion from repeated CNO flashes.


\section{CONCLUSIONS}

High-resolution, far-ultraviolet observations enabled by {\em HST} have detailed the abundances of nine metals present in the atmosphere of the extremely low-mass ($\simeq0.19$\,\msun) white dwarf \tar. Our abundance determinations are made less certain by the presence of strong rotational broadening, which we estimate to have a value of $v \sin i = 50^{+30}_{-20}$ \kms. This ELM WD is in a compact, 5.9-hr orbit with a fainter, more massive WD companion, and we show that the system exhibits both primary and secondary eclipses, indicating the binary is at high inclination as viewed from Earth ($i=86.9\pm0.4$~deg), as was originally demonstrated by \citet{VK11}. Our follow-up photometry constrains the radius of the primary to $R_1=0.093\pm0.013$~\rsun, suggesting the WD is rotating at a rate of $2.3^{+2.0}_{-1.0}$~hr.

We show that \tar\ has roughly solar abundances of Ca, P, and Ti, relative to H, and slightly sub-solar amounts of Fe and Si. We have also found that C, N, and O are all more than 100 times less abundant in this WD than the solar value. We demonstrate that the abundances are likely incompatible with rocky accretion, since the diffusion fluxes of the observed heavy elements significantly over-predict the O abundance. We extend diffusion timescale calculations to low-gravity WDs, and confirm that diffusion rates (in the absence of any radiative support) for all metals are many of orders of magnitude faster than the evolutionary WD cooling timescale. Given the increased likelihood of finding atmospheric metals in WDs with \logg\ $<5.9$, it is more likely that an internal stellar process that increases in strength with decreasing surface gravity, such as radiative levitation or rotational mixing, is counteracting diffusion.

Using an updated stellar atmosphere code, we find strong radiative support for many of the detected metals, especially Fe, P, and Si, but minimal support for Ca, which is clearly detected. Thus, radiative levitation is a plausible explanation for the presence of some of the metals in the atmosphere of this low-mass WD, but it does not fully explain the abundance patterns in \tar. An additional support mechanism, such as rotational mixing, is likely required to explain all of the observed abundances.


\section*{Acknowledgments}

We acknowledge the anonymous referee for useful comments that improved this manuscript. We also wish to acknowledge fruitful discussions with Lars Bildsten, Stephan Geier, Matteo Cantiello, and Steve Kawaler, and thank the KITP (Santa Barbara) for their kind hospitality. The research leading to these results has received funding from the European Research Council under the European Union's Seventh Framework Programme (FP/2007-2013) / ERC Grant Agreement n. 320964 (WDTracer). J.F. thanks the STFC for support in the form of an Ernest Rutherford Fellowship, and T.R.M. and E.B. thank the STFC for support under ST/L000733/1. Based on observations made with the NASA/ESA {\em HST}, obtained from the data archive at the Space Telescope Science Institute. STScI is operated by the Association of Universities for Research in Astronomy, Inc. under NASA contract NAS 5-26555. These observations are associated with {\em HST} program No. 12474, Cycle 19.


\begin{thebibliography}{}

\bibitem[Althaus et 
al.(2013)]{Althaus13} Althaus, L.~G., Miller Bertolami, M.~M., \& C{\'o}rsico, A.~H.\ 2013, \aap, 557, A19 

\bibitem[Artymowicz 
\& Lubow(1996)]{Artymowicz96} Artymowicz, P., \& Lubow, S.~H.\ 1996, \apjl, 467, L77 

\bibitem[Asplund et 
al.(2009)]{Asplund09} Asplund, M., Grevesse, N., Sauval, A.~J., \& Scott, P.\ 2009, \araa, 47, 481 

\bibitem[Barstow et al.(2014)]{Barstow14} Barstow, M.~A., 
Barstow, J.~K., Casewell, S.~L., Holberg, J.~B., 
\& Hubeny, I.\ 2014, \mnras, 544 

\bibitem[Becklin et al.(2005)]{Becklin05} Becklin, E.~E., Farihi, 
J., Jura, M., et al.\ 2005, \apjl, 632, L119 

\bibitem[Behr(2003)]{Behr03} Behr, B.~B.\ 2003, \apjs, 149, 67 

\bibitem[Bergeron et al.(1992)]{Bergeron92} Bergeron, P., Saffer, 
R.~A., \& Liebert, J.\ 1992, \apj, 394, 228 

\bibitem[Brown et al.(2013)]{BrownELMv} Brown, W.~R., Kilic, M., 
Allende Prieto, C., Gianninas, A., \& Kenyon, S.~J.\ 2013, \apj, 769, 66 

\bibitem[Charpinet et al.(1996)]{Charpinet96} Charpinet, S., 
Fontaine, G., Brassard, P., \& Dorman, B.\ 1996, \apjl, 471, L103 

\bibitem[Chayer et al.(1995)]{Chayer95} Chayer, P., Fontaine, 
G., \& Wesemael, F.\ 1995, \apjs, 99, 189

\bibitem[Chayer 
\& Dupuis(2010)]{CD10} Chayer, P., \& Dupuis, J.\ 2010, American Institute of Physics Conference Series, 1273, 394 

\bibitem[Chayer(2014)]{Chayer14} Chayer, P.\ 2014, \mnras, 437, 
L95 

\bibitem[Copperwheat et al.(2010)]{Copperwheat10} Copperwheat, C.~M., 
Marsh, T.~R., Dhillon, V.~S., et al.\ 2010, \mnras, 402, 1824 

\bibitem[Dhillon et al.(2007)]{Dhillon07} Dhillon, V.~S., Marsh, 
T.~R., Stevenson, M.~J., et al.\ 2007, \mnras, 378, 825

\bibitem[Driebe et 
al.(1999)]{Driebe99} Driebe, T., Bl{\"o}cker, T., Sch{\"o}nberner, D., \& Herwig, F.\ 1999, \aap, 350, 89 

\bibitem[Dufour et al.(2012)]{Dufour12} Dufour, P., Kilic, M., 
Fontaine, G., et al.\ 2012, \apj, 749, 6

\bibitem[Dupuis et al.(2010)]{Dupuis10} Dupuis, J., Chayer, P., 
\& H{\'e}nault-Brunet, V.\ 2010, American Institute of Physics Conference Series, 1273, 412 

\bibitem[Falcon et al.(2010)]{Falcon10} Falcon, R.~E., Winget, 
D.~E., Montgomery, M.~H., \& Williams, K.~A.\ 2010, \apj, 712, 585 

\bibitem[Farihi et al.(2009)]{Farihi09} Farihi, J., Jura, M., 
\& Zuckerman, B.\ 2009, \apj, 694, 805

\bibitem[Farihi et al.(2012)]{Farihi12} Farihi, J., 
G{\"a}nsicke, B.~T., Steele, P.~R., et al.\ 2012, \mnras, 421, 1635 

\bibitem[Farihi et al.(2013)]{Farihi13} Farihi, J., 
G{\"a}nsicke, B.~T., \& Koester, D.\ 2013, Science, 342, 218 

\bibitem[Fontaine 
\& Chayer(1997)]{Fontaine97} Fontaine, G., \& Chayer, P.\ 1997, The Third Conference on Faint Blue Stars, 169 

\bibitem[Fontaine et al.(2001)]{Fontaine01} Fontaine, G., 
Brassard, P., \& Bergeron, P.\ 2001, \pasp, 113, 409 

\bibitem[Fontaine et al.(2003)]{Fontaine03} Fontaine, G., 
Brassard, P., Charpinet, S., et al.\ 2003, \apj, 597, 518 

\bibitem[G{\"a}nsicke et al.(2006)]{Gaensicke06} G{\"a}nsicke, 
B.~T., Marsh, T.~R., Southworth, J., 
\& Rebassa-Mansergas, A.\ 2006, Science, 314, 1908 

\bibitem[G{\"a}nsicke et al.(2012)]{Gaensicke12} G{\"a}nsicke, 
B.~T., Koester, D., Farihi, J., et al.\ 2012, \mnras, 424, 333 

\bibitem[Geier(2013{\natexlab{a}})]{Geier13a} Geier, S.\ 2013a, \aap, 549, A110 

\bibitem[Geier et 
al.(2013{\natexlab{b}})]{Geier13b} Geier, S., Heber, U., Edelmann, H., et al.\ 2013b, \aap, 557, A122 

\bibitem[Gianninas et al.(2013)]{Gianninas13} Gianninas, A., 
Strickland, B.~D., Kilic, M., \& Bergeron, P.\ 2013, \apj, 766, 3 

\bibitem[Gianninas et al.(2014)]{Gianninas14} Gianninas, A., 
Hermes, J.~J., Brown, W.~R., et al.\ 2014, \apj, 781, 104 

\bibitem[Girven et al.(2012)]{Girven12} Girven, J., Brinkworth, 
C.~S., Farihi, J., et al.\ 2012, \apj, 749, 154

\bibitem[Green et al.(2012)]{Green12} Green, J.~C., Froning, 
C.~S., Osterman, S., et al.\ 2012, \apj, 744, 60 

\bibitem[Holberg et al.(1997)]{Holberg97} Holberg, J.~B., 
Barstow, M.~A., \& Green, E.~M.\ 1997, \apjl, 474, L127 

\bibitem[Hu et al.(2011)]{Hu11} Hu, H., Tout, C.~A., 
Glebbeek, E., \& Dupret, M.-A.\ 2011, \mnras, 418, 195 

\bibitem[Iben 
\& Tutukov(1984)]{Iben84} Iben, I., Jr., \& Tutukov, A.~V.\ 1984, \apjs, 54, 335 

\bibitem[Iben 
\& Tutukov(1985)]{Iben85} Iben, I., Jr., \& Tutukov, A.~V.\ 1985, \apjs, 58, 661 

\bibitem[Iben 
\& Tutukov(1986)]{IT86} Iben, I., Jr., \& Tutukov, A.~V.\ 1986, \apj, 311, 742 

\bibitem[Jeffery 
\& Saio(2006)]{Jeffery06} Jeffery, C.~S., \& Saio, H.\ 2006, \mnras, 371, 659 

\bibitem[Jura(2003)]{Jura03} Jura, M.\ 2003, \apjl, 584, L91 

\bibitem[Jura(2006)]{Jura06} Jura, M.\ 2006, \apj, 653, 613 

\bibitem[Kaplan(2010)]{Kaplan10} Kaplan, D.~L.\ 2010, \apjl, 
717, L108 

\bibitem[Kaplan et al.(2013)]{Kaplan13} Kaplan, D.~L., Bhalerao, 
V.~B., van Kerkwijk, M.~H., et al.\ 2013, \apj, 765, 158 

\bibitem[Kaplan et al.(2014)]{Kaplan14} Kaplan, D.~L., Marsh, 
T.~R., Walker, A.~N., et al.\ 2014, \apj, 780, 167 

\bibitem[Kawaler(2004)]{Kawaler04} Kawaler, S.~D.\ 2004, Stellar 
Rotation, 215, 561 

\bibitem[Koester(2009)]{Koester09} Koester, D.\ 2009, \aap, 498, 517

\bibitem[Koester(2010)]{Koester10} Koester, D.\ 2010, In Memorie della Societa Astronomica Italiana, 81, 921

\bibitem[Koester et 
al.(2014)]{Koester14} Koester, D., G{\"a}nsicke, B.~T., \& Farihi, J.\ 2014, \aap, 566, A34 

\bibitem[Klein et al.(2010)]{Klein10} Klein, B., Jura, M., 
Koester, D., Zuckerman, B., \& Melis, C.\ 2010, \apj, 709, 950 

\bibitem[Kleinman et al.(2013)]{Kleinman13} Kleinman, S.~J., 
Kepler, S.~O., Koester, D., et al.\ 2013, \apjs, 204, 5

\bibitem[Lacombe et al.(1983)]{Lacombe83} Lacombe, P., Wesemael, 
F., Fontaine, G., \& Liebert, J.\ 1983, \apj, 272, 660 

\bibitem[Marsh et al.(1995)]{Marsh95} Marsh, T.~R., Dhillon, 
V.~S., \& Duck, S.~R.\ 1995, \mnras, 275, 828

\bibitem[Marsh(2001)]{Marsh01} Marsh, T.~R.\ 2001, \mnras, 324, 
547 

\bibitem[Michaud(1980)]{Michaud80} Michaud, G.\ 1980, \aj, 85, 
589 

\bibitem[Michaud et al.(1983)]{Michaud83} Michaud, G., Vauclair, 
G., \& Vauclair, S.\ 1983, \apj, 267, 256 

\bibitem[Michaud et 
al.(2011)]{Michaud11} Michaud, G., Richer, J., \& Richard, O.\ 2011, \aap, 529, A60 

\bibitem[Paquette et al.(1986)]{Paquette86} Paquette, C., 
Pelletier, C., Fontaine, G., \& Michaud, G.\ 1986, \apjs, 61, 197 

\bibitem[Panei et al.(2007)]{Panei07} Panei, J.~A., Althaus, 
L.~G., Chen, X., \& Han, Z.\ 2007, \mnras, 382, 779 

\bibitem[Rebassa-Mansergas et al.(2011)]{RM11} 
Rebassa-Mansergas, A., Nebot G{\'o}mez-Mor{\'a}n, A., Schreiber, M.~R., 
Girven, J., G{\"a}nsicke, B.~T.\ 2011, \mnras, 413, 1121

\bibitem[Strittmatter 
\& Wickramasinghe(1971)]{SW71} Strittmatter, P.~A., \& Wickramasinghe, D.~T.\ 1971, \mnras, 152, 47

\bibitem[Talon et al.(2006)]{Talon06} Talon, S., Richard, O., 
\& Michaud, G.\ 2006, \apj, 645, 634 

\bibitem[Tremblay et al.(2011)]{Tremblay11} Tremblay, P.-E., 
Bergeron, P., \& Gianninas, A.\ 2011, \apj, 730, 128 

\bibitem[Unglaub 
\& Bues(2001)]{Unglaub01} Unglaub, K., \& Bues, I.\ 2001, \aap, 374, 570 

\bibitem[van Maanen(1917)]{vanMaanen17} van Maanen, A.\ 1917, 
\pasp, 29, 258 

\bibitem[Vauclair et al.(1978)]{Vauclair78} Vauclair, G., 
Vauclair, S., \& Michaud, G.\ 1978, \apj, 223, 920 

\bibitem[Vauclair et 
al.(1979)]{Vauclair79} Vauclair, G., Vauclair, S., \& Greenstein, J.~L.\ 1979, \aap, 80, 79 

\bibitem[Vennes et al.(2011)]{VK11} Vennes, S., Thorstensen, J.~R., Kawka, A., et al.\ 2011, \apjl, 737, L16 

\bibitem[Veras(2014)]{Veras14} Veras, D.\ 2014, Celestial 
Mechanics and Dynamical Astronomy, 10 

\bibitem[von Hippel et al.(2007)]{vonHippel07} von Hippel, T., 
Kuchner, M.~J., Kilic, M., Mullally, F., 
\& Reach, W.~T.\ 2007, \apj, 662, 544 

\bibitem[Weidemann(1960)]{Weidemann60} Weidemann, V.\ 1960, \apj, 
131, 638 

\bibitem[Xu et al.(2013)]{Xu13} Xu, S., Jura, M., Klein, B., 
Koester, D., \& Zuckerman, B.\ 2013, \apj, 766, 132 

\bibitem[Zuckerman et al.(2003)]{Zuckerman03} Zuckerman, B., 
Koester, D., Reid, I.~N., H{\"u}nsch, M.\ 2003, \apj, 596, 477 

\bibitem[Zuckerman et al.(2007)]{Zuckerman07} Zuckerman, B., 
Koester, D., Melis, C., Hansen, B.~M., \& Jura, M.\ 2007, \apj, 671, 872 

\bibitem[Zuckerman et al.(2010)]{Zuckerman10} Zuckerman, B., Melis, 
C., Klein, B., Koester, D., \& Jura, M.\ 2010, \apj, 722, 725 

\end{thebibliography}
\end{document}